\documentclass[
 aps,
 prc,
 reprint,
 superscriptaddress,
 amsmath,amssymb,
]{revtex4-2}

\usepackage[dvipdfmx]{graphicx}  % Include figure files
\usepackage{dcolumn}  % Align table columns on decimal point

\usepackage{bm}  % Bold math

\usepackage[
 bookmarks=false
]{hyperref}  % Add hypertext capabilities

\usepackage{physics}

\usepackage{color}

\hypersetup{
 pdfpagemode={UseOutlines},
 bookmarksopen=true,
 bookmarksopenlevel=0,
 hypertexnames=false,
 colorlinks=true,  % Set to false to disable coloring links
 citecolor=blue,   % The color of citations
 linkcolor=blue,   % The color of references to document elements (sections, figures, etc)
 urlcolor=blue,    % The color of hyperlinks (URLs)
 allcolors=blue,   % The color of all
 pdfstartview={FitV},
 unicode,
 breaklinks=true,
}

\usepackage{ulem}

\usepackage{xcolor}

\newcommand{\trA}{\textrm{A}}
\newcommand{\trB}{\textrm{B}}
\newcommand{\trL}{\textrm{L}}

\newcommand{\mathT}{\mathcal{T}}

\newcommand{\vcK}{\vb*{K}}
\newcommand{\vcR}{\vb*{R}}

\newcommand{\vcq}{\vb*{q}}

\newcommand{\vkap}{\vb*{\kappa}}

\newcommand{\nn}{\nonumber\\}

\begin{document}

%---- Title of paper ----%
\title{Maris polarization in the (\textit{p},\textit{pd}) reaction}

%---- Author ----%
\author{Yoshiki Chazono}
\email[]{chazono.yoshiki.907@m.kyushu-u.ac.jp}
\affiliation{Department of Physics, Kyushu University, Fukuoka 819-0395, Japan}
%\altaffiliation{}  % Alternative affiliation

%---- Date ----%
\date{\today}

%---- Abstract ----%
\begin{abstract}
\begin{description}
\item[Background]
Proton-induced knockout reactions at intermediate energies provide a clean probe for nuclear clusters with relatively small theoretical uncertainties.
The Maris polarization, which is the effective polarization of a particle inside a nucleus arising from nuclear absorption and spin-orbit coupling, has been used in proton knockout to determine the total angular momentum $j$ of the removed protons.
However, its manifestation in cluster knockout remains unexplored.
\item[Purpose]
We theoretically demonstrated that the Maris polarization can be observed via the vector analyzing power $A_y$ of the proton-induced deuteron knockout ($p,pd$) reaction in imbalanced kinematics.
\item[Method]
First, we computed the spin correlation coefficient $C_{y,y}$ of $p$-$d$ elastic scattering, which is an elementary process, to identify suitable kinematics for the Maris polarization.
Subsequently, we calculated the $A_y$ values of the ($p,pd$) reaction at $250$~MeV in forward kinematics for deuteron-cluster orbits with $j = 1$, $2$, and $3$.
\item[Result]
The large positive $C_{y,y}$ values at $p$-$d$ scattering angles of $\sim 40^\circ$ are consistent with the experimental data.
In the corresponding ($p,pd$) kinematics, the signs of $A_y$ for $j = 3$ and $1$ orbits are positive and negative, respectively, indicating effective upward and downward polarizations of the deuterons in the nucleus.
The $A_y$ value for $j = 2$ orbit lies between those for the other two orbits, which can also be explained by the Maris polarization, with the deuteron being knocked out from regions near the poles of the $y$-axis.
These results are nearly independent of the internal state of the deuteron and the nucleon-nucleon effective interactions used to describe the elementary process.
\item[Conclusion]
We theoretically demonstrated that the Maris polarization occurs in the ($p,pd$) reaction under imbalanced kinematics.
This work may provide information on the orbital motion of the deuteron cluster in a nucleus and could lead to the establishment of the concept of deuteron-cluster orbit.
Further experimental and theoretical studies are required to improve the quantitative understanding of this effect.
\end{description}
\end{abstract}

%---- Main text ----%
\maketitle

%----
\section{Introduction\label{sec:introduction}}
Although the independent-particle picture, wherein nucleons move nearly independently in a nucleus, is well established, several nucleons can form clusters that behave as single entities.
Nuclear clustering is a prominent manifestation of nonuniformity in nuclei and is relevant to, for example, $\alpha$ decay in heavy nuclei and the synthesis of elements in the universe~\cite{TUesaka24a}.
Consequently, this subject has been extensively researched, both experimentally and theoretically.

Proton-induced knockout reactions provide a direct means to observe clusters and their motion in nuclei~\cite{CSamanta82, KazukiYoshida16, KazukiYoshida18, KazukiYoshida19, JTanaka21, KazukiYoshida22, YChazono21, YChazono22, TEdagawa23, TUesaka24, CSLee26, RMatsumura26, RTsuji26}.
In such reactions, a proton of several hundred MeV per nucleon collides with a nucleus and removes a constituent particle from it.
Owing to the high incident energy, the disturbance of the residual system by the proton is small, and the reaction can be approximately described as scattering between the proton and the particle to be removed~\cite{GJacob66, GJacob73, NSChant77, NSChant83, TWakasa17, KOgata24}.
This reduces the theoretical uncertainties involved in extracting cluster information compared with those in other reactions.
Using this approach, the ONOKORO project has investigated the presence of light clusters---deuterons, tritons, $^3$He, and $\alpha$---in stable and unstable nuclei over a wide mass range~\cite{TUesaka24, TUesaka24a}.
In addition, the proton-induced deuteron knockout ($p,pd$) reaction may be useful for exploring the three-nucleon force effects in the nuclear medium because the reaction can be regarded as proton-deuteron elastic scattering inside nuclei.

Knockout data provide the single-particle properties of the particle to be removed, which are characterized by radial quantum number $n$, orbital angular momentum $\ell$, and total angular momentum $j$.
The shape of the knockout cross section strongly depends on $\ell$, and the value of $\ell$ for the knocked-out particle in the nucleus can be determined by comparing the calculated cross section with the experimental data.
By contrast, the cross sections for different values of $j$ coupled with the same $\ell$ exhibit nearly identical shapes.
Therefore, another indicator is needed to determine the $j$ value.
The Maris polarization---an effective polarization of the particle to be knocked out that arises from the strong nuclear absorption of a low-energy ejectile combined with spin-orbit coupling~\cite{TWakasa17, Shubhchintak18}---is a good indicator for determining $j$, and it was originally proposed by Jacob and Maris~\cite{GJacob73a, GJacob76}.
Using the strong spin correlation of the nucleon-nucleon elastic cross section at intermediate energies, this effect was observed in the proton-induced proton knockout ($p,2p$) reaction through the reversal of the signs of the vector analyzing powers $A_y$, for example, between the $p_{3/2}$ and $p_{1/2}$ orbits in $^{16}$O~\cite{PKitching80, PKitchingB85}.

Because of the generality of the mechanism, the Maris polarization is also expected to occur in cluster knockout reactions; however, this has rarely been discussed.
Among the aforementioned light clusters, the deuteron, owing to its spin $1$, is expected to exhibit the Maris polarization with characteristics different from those observed in nucleon knockout reactions.
Experimentally, three peaks corresponding to the $j = \ell + 1$, $\ell$, and $\ell - 1$ states of the residual nucleus will be observed in the energy spectrum of the ($p,pd$) reaction, and the value of $\ell$ and $j$ can be determined from the cross section shape and the Maris polarization, respectively, which could lead to the establishment of the concept of deuteron-cluster orbit.
Therefore, the Maris polarization was considered in the ($p,pd$) reaction, and we numerically demonstrated it in this study.

The remainder of this paper is organized as follows.
Section~\ref{sec:Framework} describes the ($p,pd$) reaction within the distorted wave impulse approximation (DWIA) and briefly reviews the Maris polarization.
In addition, the spin correlation coefficient $C_{y,y}$ of the elementary $p$-$d$ process and the vector analyzing power $A_y$ of the ($p,pd$) reaction are introduced.
Section~\ref{sec:Result} presents numerical results and discussion.
Finally, Sec.~\ref{sec:Summary} provides a summary and a perspective.
%~\\

%----
\section{Theoretical framework\label{sec:Framework}}
The deuteron is a fragile nucleus and its breakup effects would change ($p,pd$) cross sections by up to approximately $20\%$ at high incident-proton energies, as discussed in Ref.~\cite{YChazono22}.
However, this difference is expected to be relatively small in this study because $A_y$ is defined as the ratio of cross sections, as mentioned in Sec.~\ref{sec:Ay}; that is, contributions of these effects are expected to be reduced between the numerator and denominator.
Therefore, we describe the ($p,pd$) reaction in forward kinematics within the DWIA framework; a quantitative treatment of spin-dependent breakup effects remains a subject for future work.
The deuteron in the target nucleus is assumed to be the same as that in free space in the following formalism; this assumption is discussed in Sec.~\ref{sec:NuclMedEff}.
The incident proton, emitted proton, and knocked-out deuteron are labeled as particles 0, 1, and 2, respectively.
The target (residual) nucleus is denoted as A (B) with mass number $A$ ($B$).
For particles or nuclei $i$ ($i = 0, 1, 2, \trA, \trB$), the total energy, kinetic energy, and asymptotic momentum (in units of $\hbar$) are denoted as $E_i$, $T_i$, and $\vcK_i$, respectively.
Quantities with superscript L are evaluated in the laboratory (L) frame, whereas those without are evaluated in the center-of-mass (c.m.) frame of the $p$-A system.
Although the Coulomb interaction and the spin-orbit terms of the optical potentials are not explicitly presented in the formalism for simplicity, these are included in the numerical calculation using a standard method.
The $z$-axis is oriented along the beam direction, and we assume coplanar kinematics; that is, all momenta lie in the plane parallel to the $z$-$x$ plane.
%~\\

%----
\subsection{(\textit{p},\textit{pd}) reaction within DWIA\label{sec:DWIA}}
The DWIA transition matrix of the ($p,pd$) reaction is given by
\begin{align}
T_{\mu_1 \mu_2 \mu_0 \mu_j}
&=
\sum_{\mu_d} \tilde{t}_{pd, \mu_1 \mu_2 \mu_0 \mu_d} (\vkap', \vkap) \nn
&\quad \times
\int \dd{\vcR} D_{\vcK_1, \vcK_2, \vcK_0} (\vcR) \nn
&\quad \times
\sum_m (\ell m 1 \mu_d | j \mu_j) \varphi_{n \ell j m} (\vcR),
\label{eq:Tmat_ppd}
\end{align}
where $\tilde{t}_{pd, \mu_1 \mu_2 \mu_0 \mu_d}$ is the transition amplitude of the elementary $p$-$d$ process, and $\vkap$ and $\vkap'$ are the $p$-$d$ relative momenta in the initial and final states, respectively.
The characters $\mu_i$ and $\mu_d$ are the spin projections of particle $i$ and the deuteron in nucleus A, respectively.
The function $D_{\vcK_1, \vcK_2, \vcK_0}$ is defined as
\begin{align}
&D_{\vcK_1, \vcK_2, \vcK_0} (\vcR) \nn
&\quad \equiv
\chi_{\vcK_1}^{(-) *} (\vcR) \chi_{\vcK_2}^{(-) *} (\vcR)
\chi_{\vcK_0}^{(+)} (\vcR) \mathrm{e}^{- 2 i \vcK_0 \cdot \vcR / A},
\label{eq:ChiProduct}
\end{align}
where $\chi_{\vcK_i}$ denotes the distorted wave of particle $i$, and the superscripts $(+)$ and $(-)$ indicate that $\chi_{\vcK_i}$ satisfy the outgoing and incoming boundary conditions, respectively.
This $D_{\vcK_1, \vcK_2, \vcK_0}$ represents the suppression of the reaction at $\vcR$ owing to absorption.
$\varphi_{n \ell j m}$ denotes the deuteron-cluster wave function bound to nucleus B, and $(\ell m 1 \mu_d | j \mu_j)$ is a Clebsch-Gordan coefficient.
The characters $m$ and $\mu_j$ denote the third components of $\ell$ and $j$, respectively.

To obtain Eq.~\eqref{eq:ChiProduct}, we used the asymptotic momentum approximation (AMA)~\cite{YWatanabe99, KazukiYoshida16, TWakasa17} in which the short-range propagation from $\vcR$ to $\vcR + \Delta \vcR$ is described by a plane wave with $\vcK_i$, that is,
\begin{align}
\chi_{\vcK_i} (\vcR + \Delta \vcR) \approx
\chi_{\vcK_i} (\vcR) \mathrm{e}^{i \vcK_i \cdot \Delta \vcR}.
%\label{eq:}
\end{align}
Under this approximation, the selection of one kinematics of the ($p,pd$) reaction determined the corresponding kinematics of the elementary $p$-$d$ process.
The validity of AMA in the ($p,pd$) reaction was investigated in Ref.~\cite{KazukiYoshida24}, and it was demonstrated that AMA works well at high incident energies.

Using Eq.~\eqref{eq:Tmat_ppd}, the triple-differential cross section (TDX) of the ($p,pd$) reaction is expressed as
\begin{align}
\frac{\dd[3]{\sigma^\trL}}{\dd{E}_1^\trL \dd{\Omega}_1^\trL \dd{\Omega}_2^\trL}
&=
F_\textrm{kin}^\trL \frac{E_1 E_2 E_\trB}{E_1^\trL E_2^\trL E_\trB^\trL}
\frac{(2 \pi)^4}{\hbar v_\alpha^\trL} \frac{1}{2 (2 j + 1)} \nn
&\quad \times
\sum_{\mu_1 \mu_2 \mu_0 \mu_j} \abs{T_{\mu_1 \mu_2 \mu_0 \mu_j}}^2,
\label{eq:TDX_ppd}
\end{align}
where the kinematical factor $F_\textrm{kin}^\trL$ is defined as
\begin{align}
F_\textrm{kin}^\trL \equiv
\frac{E_1^\trL K_1^\trL E_2^\trL K_2^\trL}{(\hbar c)^4}
\qty[1 + \frac{E_2^\trL}{E_\trB^\trL}
- \frac{E_2^\trL}{E_\trB^\trL} \frac{\vcq^\trL \cdot \vcK_2^\trL}{(K_2^\trL)^2}]^{-1}.
%\label{eq:}
\end{align}
Here, $\vcq^\trL \equiv \vcK_0^\trL - \vcK_1^\trL$ is the momentum transfer, and $v_\alpha^\trL$ is the relative speed between particle 0 and nucleus A.
% \vcq^\trL \equiv \vcK_0^\trL + \vcK_\trA^\trL - \vcK_1^\trL
%~\\

%----
\subsection{Maris polarization\label{sec:MarisPol}}

%\vspace{-0.5cm}
\begin{figure}[htpb]
 \centering
 \includegraphics[width=0.40\textwidth]{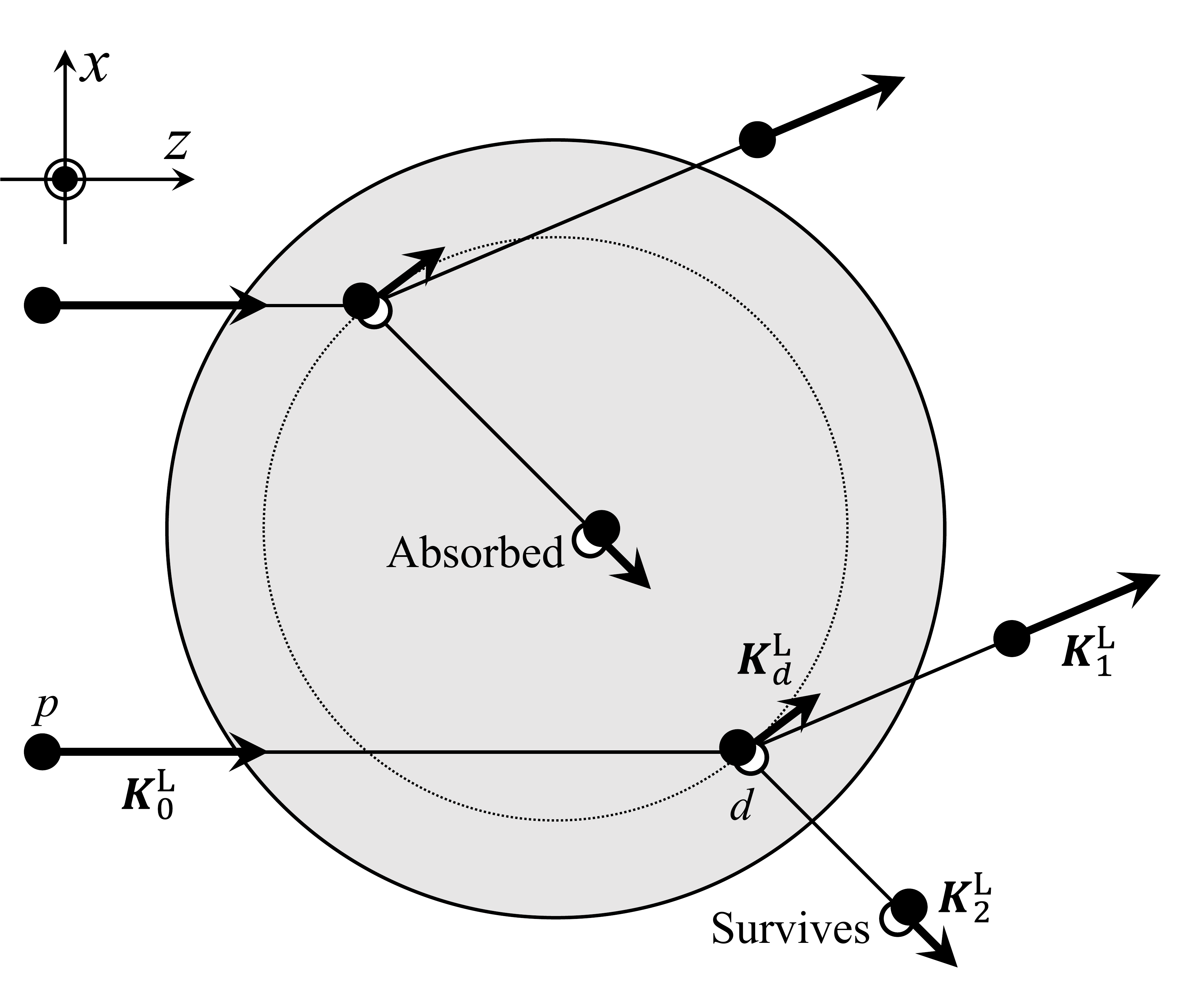}
% \vspace{-0.3cm}
 \caption{Schematic of the Maris polarization.
 \label{fig:MarisPol}}
\end{figure}

In this section, we briefly review the Maris polarization~\cite{GJacob73a, GJacob76, PKitching80, PKitchingB85, TWakasa17, Shubhchintak18}.
We consider the kinematics illustrated in Fig.~\ref{fig:MarisPol}, where $T_2^\trL$ is significantly lower than $T_1^\trL$.
As mentioned previously, all momenta are assumed to lie on the $z$-$x$ plane in this figure.
The deuteron scattered at the upper side of Fig.~\ref{fig:MarisPol} must traverse the nuclear medium to exit, in contrast to that originating from the lower side.
Because the $T_2^\trL \ll T_1^\trL$ condition is considered, the absorption effects on particle $2$ are expected to be significantly stronger than those on particle $1$.
Consequently, under such imbalanced kinematics, the deuteron knocked out from the lower side of the nucleus predominantly contributes to the ($p,pd$) reaction.
If the $z$ component of the momentum of the internal deuteron $K_{d,z}^\trL$, as estimated from momentum conservation, is positive (Fig.~\ref{fig:MarisPol}), it moves counterclockwise in orbit with a given $\ell$ in a classical picture.
Considering spin-orbit coupling, three cluster orbits with $j = \ell + 1$, $\ell$, and $\ell - 1$ are relevant; these correspond to deuteron spins that are parallel, ``orthogonal'', and antiparallel to $\ell$, respectively.
Therefore, the combination of absorption and spin-orbit coupling generates an effective polarization of the knocked-out deuteron in nucleus A, which is referred to as the Maris polarization.
%~\\

%----
\subsection{Spin correlation coefficient $C_{y,y}$\label{sec:Cyy}}
In the ($p,2p$) reaction, the fact that the spin-triplet $p$-$p$ elastic cross section considerably exceeds the spin-singlet cross section at intermediate energies is exploited to observe this effect.
If the elementary $p$-$d$ process exhibits a similar spin preference, the Maris polarization can be observed in the ($p,pd$) reaction.
The spin correlation coefficient $C_{y,y}$ quantifies whether spin-parallel or spin-antiparallel scattering dominates the $p$-$d$ elastic scattering.

For $p$-$d$ elastic scattering, $C_{y,y}$ is defined as
\begin{align}
C_{y,y} =
\frac{1}{N_{pd}}
\Tr[\tilde{\vb*{t}}_{pd} \vb*{\tau}_y \vb*{\sigma}_y \tilde{\vb*{t}}_{pd}^\dagger],
\label{eq:Cyy_pd}
\end{align}
where $\tilde{\vb*{t}}_{pd}$ is the $6 \times 6$ transition matrix with elements $\tilde{t}_{pd, \mu_1 \mu_2 \mu_0 \mu_d}$.
The denominator $N_{pd}$ is calculated as
\begin{align}
N_{pd} =
\Tr[\tilde{\vb*{t}}_{pd} \tilde{\vb*{t}}_{pd}^\dagger] =
\sum_{\mu_1 \mu_2 \mu_0 \mu_d} \abs{\tilde{t}_{pd, \mu_1 \mu_2 \mu_0 \mu_d}}^2,
\label{eq:N_pd}
\end{align}
which is the $p$-$d$ elastic cross section without the mass factor. The matrices $\vb*{\tau}_y$ and $\vb*{\sigma}_y$ are the $y$-components of the $6 \times 6$ deuteron and proton spin matrices, respectively.
In the case of quantization along the $y$-axis, which is expressed by the superscript $(y)$, these matrices are denoted by
\begin{align}
\vb*{\tau}_y^{(y)} =
\qty(\mqty{
1 & 0 & 0 & 0 &  0 &  0 \\ 0 & 1 & 0 & 0 &  0 &  0 \\
0 & 0 & 0 & 0 &  0 &  0 \\ 0 & 0 & 0 & 0 &  0 &  0 \\
0 & 0 & 0 & 0 & -1 &  0 \\ 0 & 0 & 0 & 0 &  0 & -1})
\label{eq:tau_y}
\end{align}
and
\begin{align}
\vb*{\sigma}_y^{(y)} =
\qty(\mqty{
1 & 0 & 0 & 0 & 0 & 0 \\ 0 & -1 & 0 &  0 & 0 &  0 \\
0 & 0 & 1 & 0 & 0 & 0 \\ 0 &  0 & 0 & -1 & 0 &  0 \\
0 & 0 & 0 & 0 & 1 & 0 \\ 0 &  0 & 0 &  0 & 0 & -1}).
\label{eq:sig_y}
\end{align}
Using these, $C_{y,y}$ can be rewritten as
\begin{align}
C_{y,y} =
\bar{N}_{pd, ++}^{(y)} + \bar{N}_{pd, --}^{(y)}
- \bar{N}_{pd, +-}^{(y)} - \bar{N}_{pd, -+}^{(y)}.
\label{eq:Cyy_pd_y}
\end{align}
The first and second terms in Eq.~\eqref{eq:Cyy_pd_y} are defined as follows:
\begin{align}
\bar{N}_{pd, \pm\pm}^{(y)} &=
\frac{1}{N_{pd}}
\sum_{\mu_1 \mu_2} \abs{\tilde{t}_{pd, \mu_1 \mu_2, \pm 1/2, \pm 1}^{(y)}}^2.
\label{eq:N_pd_pmpm}
\end{align}
These indicate how the spin-parallel processes contribute to $p$-$d$ elastic scattering, whereas the third and fourth terms,
\begin{align}
\bar{N}_{pd, \pm\mp}^{(y)} &=
\frac{1}{N_{pd}}
\sum_{\mu_1 \mu_2} \abs{\tilde{t}_{pd, \mu_1 \mu_2, \pm 1/2, \mp 1}^{(y)}}^2,
\label{eq:N_pd_pmmp}
\end{align}
represent the contributions of the spin-antiparallel processes.
Thus, $C_{y,y} > 0$ indicates the dominance of spin-parallel over spin-antiparallel processes in $p$-$d$ elastic scattering, and vice versa.
Therefore, kinematics with a large positive $C_{y,y}$ are favorable for observing the Maris polarization.
%~\\

%----
\subsection{Vector analyzing power $A_y$\label{sec:Ay}}
The vector analyzing power $A_y$ is suitable for measuring the Maris polarization, which is defined as
\begin{align}
A_y =
\frac{\sum_{\mu_2 \mu_j} \Tr[\mathT_{\mu_2 \mu_j} \sigma_y \mathT_{\mu_2 \mu_j}^\dagger]
}{\sum_{\mu_2 \mu_j} \Tr[\mathT_{\mu_2 \mu_j} \mathT_{\mu_2 \mu_j}^\dagger]},
\label{eq:Ay_ppd}
\end{align}
with
\begin{align}
\mathT_{\mu_2 \mu_j} =
\qty(\mqty{
T_{ 1/2, \mu_2, 1/2, \mu_j} & T_{ 1/2, \mu_2, -1/2, \mu_j} \\
T_{-1/2, \mu_2, 1/2, \mu_j} & T_{-1/2, \mu_2, -1/2, \mu_j}}).
\label{eq:Tmat_ppd_22}
\end{align}
Here, $\sigma_y$ denotes the $y$-component of the Pauli matrix.
Equation~\eqref{eq:Ay_ppd} assumes that only a certain cluster orbit contributes to the ($p,pd$) reaction, and summations over $\ell$ and $j$ are required in both the numerator and denominator when certain orbits are related to the reaction.
By selecting the $y$-axis for quantization, Eq.~\eqref{eq:Ay_ppd} is reduced to
\begin{align}
A_y =
\frac{\dd{\sigma_+^{(y)}} - \dd{\sigma_-^{(y)}}
}{\dd{\sigma_+^{(y)}} + \dd{\sigma_-^{(y)}}},
\label{eq:Ay_ppd_y}
\end{align}
where $\dd{\sigma_+^{(y)}}$ ($\dd{\sigma_-^{(y)}}$) denotes the TDX obtained using spin-up (spin-down) incident protons.
A positive (negative) $A_y$ implies that the reaction is predominantly induced by spin-up (spin-down) particle $0$.
For kinematics with positive $C_{y,y}$, a positive (negative) $A_y$ can be interpreted as the effective upward (downward) polarization of the deuteron inside nucleus A.
%~\\

%----
\section{Results and discussions\label{sec:Result}}

%----
\subsection{Input\label{sec:Input}}

%\vspace{-0.5cm}
\begin{figure}[htpb]
 \centering
 \includegraphics[width=0.40\textwidth]{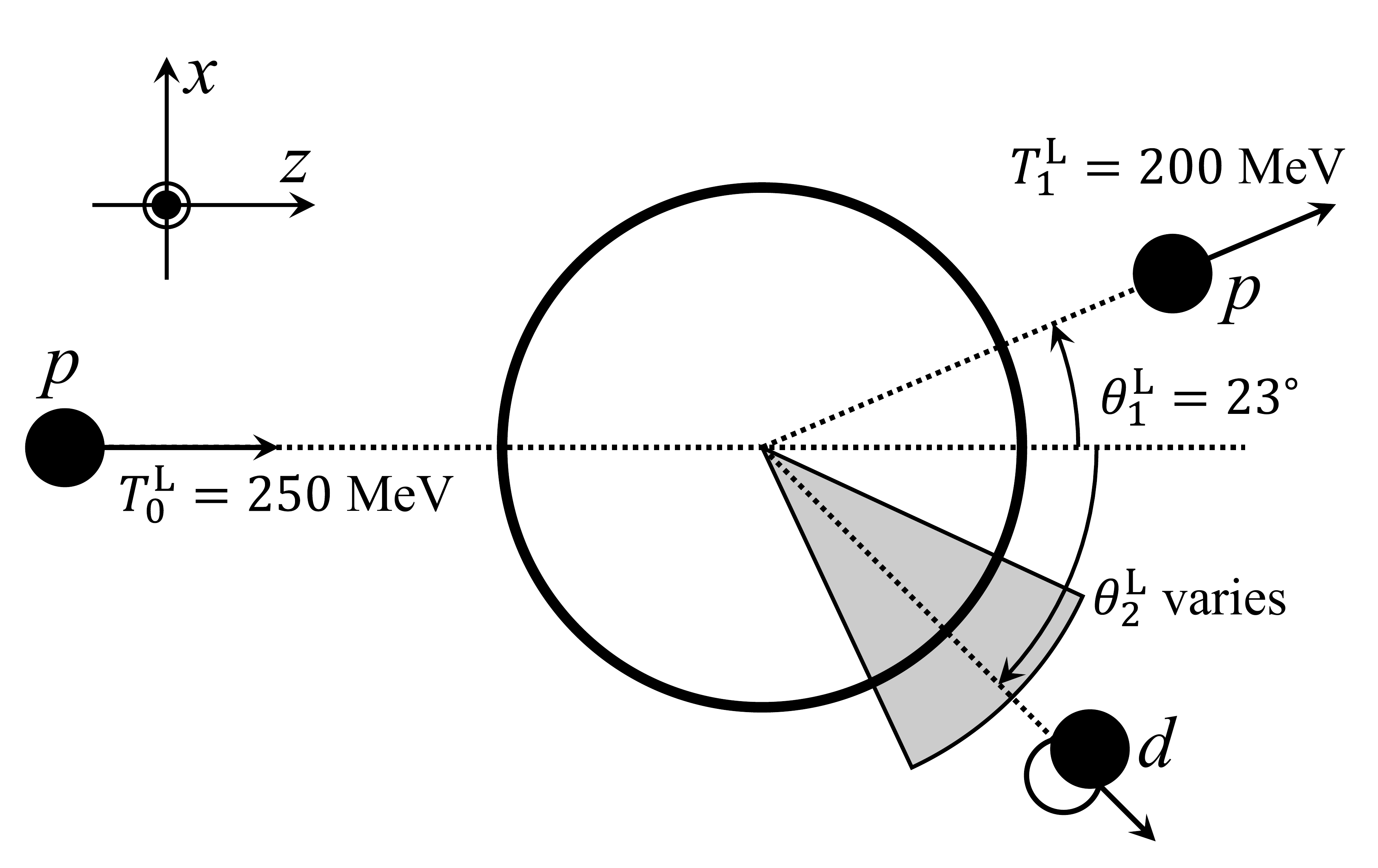}
% \vspace{-0.3cm}
 \caption{Kinematics of the $^{56}$Ni($p,pd$)$^{54}$Co$^*$ reaction.
 \label{fig:Kinematics}}
\end{figure}

We calculated the $^{56}$Ni($p,pd$)$^{54}$Co$^*$ reaction at $250$~MeV using the computer code \textsc{pikoe}~\cite{KOgata24} with expansions for the ($p,pd$) reaction.
The target nucleus is in its ground state, and the residual nucleus is considered to be in the $0.94$~MeV excited state.
%As illustrated in Fig.~\ref{fig:Kinematics}, we set the kinetic energy $T_1^\trL$ and the emission (polar) angle $\theta_1^\trL$ to $199$~MeV and $28^\circ$, respectively, and vary $\theta_2^\trL$ from $25^\circ$ to $65^\circ$; $T_2^\trL$ remains nearly constant at $\sim 30$~MeV over this range.  % Maris1
As illustrated in Fig.~\ref{fig:Kinematics}, we set the kinetic energy $T_1^\trL$ and the emission (polar) angle $\theta_1^\trL$ to $200$~MeV and $23^\circ$, respectively, and varied $\theta_2^\trL$ from $25^\circ$ to $65^\circ$; $T_2^\trL$ remained nearly constant at $\sim 30$~MeV over this range.  % Maris2
This setup corresponds to an elementary process with large positive $C_{y,y}$, as discussed in Sec.~\ref{sec:Cyy_pd}.

The elementary $p$-$d$ amplitude $\tilde{t}_{pd, \mu_1 \mu_2 \mu_0 \mu_d}$ was calculated from the Melbourne $g$-matrix interaction in free space~\cite{KAmosB00}.
This method has been used in Ref.~\cite{YChazono22} to describe the ($p,pd$) reaction, and a comparison with experimental data was performed.
The calculation reasonably reproduced the data at certain energies.
Details are available in Ref.~\cite{YChazono22} and the references therein.

The optical potentials with the EDAD1 parameter set~\cite{SHama90, EDCooper93, EDCooper09} of the Dirac phenomenology were employed as distorting potentials for the $p$-A and $p$-B systems, assuming a uniformly charged sphere with the radius proposed in Ref.~\cite{AJKoning03}.
The wave functions of protons were multiplied by the Darwin factor, which stems from the reduction of the Dirac equation to a Schr\"{o}dinger-like one~\cite{LGArnold81, SHama90}, for the relativistic correction.
For the $d$-B system, the global deuteron optical potential parameterized by An and Cai~\cite{HAn06} was used.
The nonlocality correction to the deuteron distorted wave was included via the Perey factor~\cite{FPerey62} with the range parameter $\beta = 0.54$~fm~\cite{MIgarashiM77}.

For the deuteron-cluster orbits in nucleus A, both $n$ and $\ell$ were set to $2$, and three $j$ values, $1$, $2$, and $3$ were considered.
These orbits are denoted by $2D_j$.
The binding potential generating $\varphi_{n \ell j m}$ is expressed as
\begin{align}
V^\textrm{Bind} (R) =
V^\textrm{CE} (R) + V^\textrm{SO} (R) \vb*{\ell} \cdot \vb*{S} + V^\textrm{Coul} (R).
\label{eq:BoundPot}
\end{align}
The central term $V^\textrm{CE}$ and spin-orbit term $V^\textrm{SO}$ are
\begin{align}
V^\textrm{CE} (R) =
- V_0^\textrm{CE} f (R, R^\textrm{CE}, a^\textrm{CE})
\label{eq:BoundPotCE}
\end{align}
and
\begin{align}
V^\textrm{SO} (R) =
- V_0^\textrm{SO} \frac{2}{R}  % \lambda_\pi^2 = 2
\dv{R} f (R, R^\textrm{SO}, a^\textrm{SO}),
\label{eq:BoundPotSO}
\end{align}
respectively, where the function $f$ has the Woods-Saxon form:
\begin{align}
f (R, R^\textrm{X}, a^\textrm{X}) =
\frac{1}{1 + \exp[(R - R^\textrm{X}) / a^\textrm{X}]},
\label{eq:WoodsSaxonPot}
\end{align}
with $\textrm{X} =$ CE and SO.
We set the radial parameter to $R^\textrm{X} = 1.41 \times B^{1/3}$~fm and the diffuseness to $a^\textrm{X} = 0.65$~fm~\cite{CSamanta82}.
The depth of the central term $V_0^\textrm{CE}$ was adjusted to reproduce the deuteron separation energy $19.99$~MeV, whereas that of the spin-orbit term $V_0^\textrm{SO}$ was maintained at $10.00$~MeV.
The Coulomb term $V^\textrm{Coul}$ used a uniformly charged sphere with a Coulomb radius of $1.41 \times B^{1/3}$~fm.
%~\\

%----
\subsection{$C_{y,y}$ of elementary process\label{sec:Cyy_pd}}

\begin{figure}[htpb]
 \centering
 \includegraphics[width=0.45\textwidth]{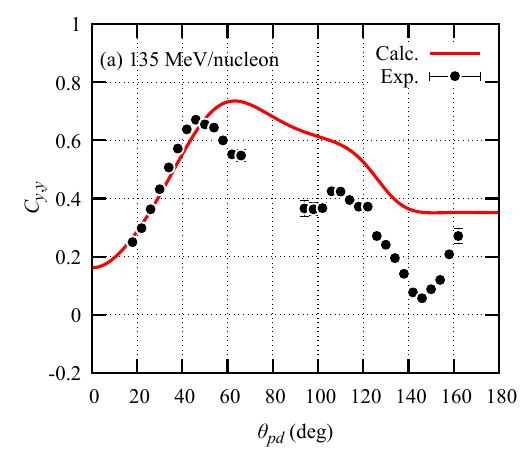}
 \includegraphics[width=0.45\textwidth]{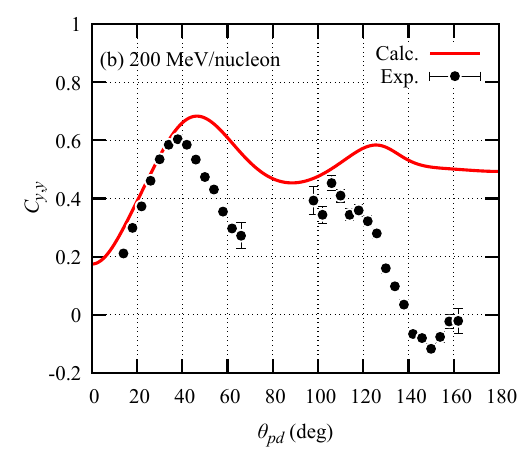}
% \vspace{-0.3cm}
 \caption{Angular distribution of $C_{y,y}$ of the $p$-$d$ elastic scattering at (a) $135$ and (b) $200$~MeV/nucleon.
 The scattering angle $\theta_{pd}$ is defined in the c.m.~frame of the $p$-$d$ system.
 In each panel, the numerical result (solid line) is compared with the experimental data (filled circles)~\cite{BvPrzewoski06}.
 \label{fig:Cyy_pd}}
\end{figure}

Figure~\ref{fig:Cyy_pd} shows $C_{y,y}$ of $p$-$d$ elastic scattering as a function of the scattering angle $\theta_{pd}$ in the $p$-$d$ c.m.~frame.
The solid lines are the results calculated at $p$-$d$ scattering energies $T_{pd}$ of (a) $135$ and (b) $200$~MeV/nucleon using Eq.~\eqref{eq:Cyy_pd_y}, which are compared with the experimental data~\cite{BvPrzewoski06} represented by filled circles.
In each case, the experimental data have a large positive value around $\theta_{pd} \approx 40^\circ$, and their height and position indicate weak energy dependence.
Although the calculations begin to overestimate the experimental data at approximately (a) $50^\circ$ and (b) $45^\circ$, they reproduce the data well for $\theta_{pd} \lesssim 40^\circ$.
The setup described in Sec.~\ref{sec:Input}, $\theta_2^\trL = 25^\circ$, $45^\circ$, and $65^\circ$ correspond to $\theta_{pd} \approx 45^\circ$, $40^\circ$, and $35^\circ$, respectively, whereas the values of $T_{pd}$ are approximately $140$, $175$, and $210$~MeV/nucleon.
Experimental data covering a wide range of angles are limited; therefore, we cannot directly compare the calculation with experimental data at $T_{pd} = 175$~MeV/nucleon.
However, because of the weak energy dependence of $C_{y,y}$, these conditions would be favorable for observing the Maris polarization, as discussed in Sec.~\ref{sec:Cyy}.
%~\\

%----
\subsection{$A_y$ of $^{56}$Ni(\textit{p},\textit{pd})$^{54}$Co reaction\label{sec:56Nippd54Co}}

%\begin{figure}[htpb]
% \centering
%% \includegraphics[width=0.45\textwidth]{Fig_TDX_56Nippd54Co_250MeV_Maris1.pdf}
% \includegraphics[width=0.45\textwidth]{Fig_TDX_56Nippd54Co_250MeV_Maris2.pdf}
% \caption{TDX of the $^{56}$Ni($p,pd$)$^{54}$Co$^*$ reaction at $250$~MeV as a function of $\theta_2^\trL$.
% The solid, dashed, and dotted lines are the calculated results where the deuterons to be knocked-out move in the $2D_3$, $2D_2$, and $2D_1$ orbits, respectively.
% \label{fig:TDX_ppd}}
%\end{figure}

%\vspace{-0.5cm}
\begin{figure}[htpb]
 \centering
 \includegraphics[width=0.45\textwidth]{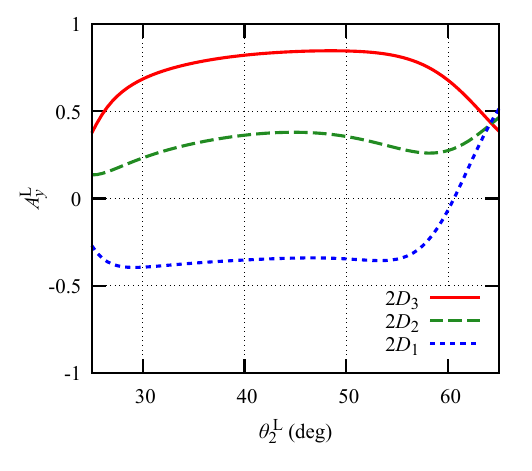}
% \vspace{-0.3cm}
 \caption{$A_y$ of the $^{56}$Ni($p,pd$)$^{54}$Co$^*$ reaction at $250$~MeV as a function of $\theta_2^\trL$.
 The solid, dashed, and dotted lines represent the calculated results in which the deuterons to be knocked out move in orbits $2D_3$, $2D_2$, and $2D_1$, respectively.
 \label{fig:Ay_ppd}}
\end{figure}

Figure~\ref{fig:Ay_ppd} shows the $\theta_2^\trL$ dependence of $A_y$ for the $^{56}$Ni($p,pd$)$^{54}$Co$^*$ reaction at $250$~MeV.
The solid, dashed, and dotted lines correspond to the numerical results of deuterons to be knocked out from orbits $2D_3$, $2D_2$, and $2D_1$, respectively.
The solid and dotted lines have opposite signs over nearly the entire angular range, and the dashed line lies between them.
As the solid (dotted) line is positive (negative), the deuteron in orbit $2D_3$ ($2D_1$) is preferentially knocked out by a spin-up (spin-down) incident proton.
According to momentum conservation, the internal deuteron in nucleus A moves to the right in the considered kinematics, as shown in Fig.~\ref{fig:MarisPol} ($K_{d,z}^\trL > 0$), which corresponds to counterclockwise motion.
Therefore, the deuterons in orbits $2D_3$ and $2D_1$ are effectively polarized upward and downward, respectively, which is consistent with the mechanism explained in Sec.~\ref{sec:MarisPol}, and Fig.~\ref{fig:Ay_ppd} shows that the Maris polarization can be observed in ($p,pd$) under imbalanced kinematics.
As the calculated $C_{y,y}$ of the elementary $p$-$d$ process deviates from the data at $\theta_{pd} \gtrsim 45^\circ$, the predictions shown in Fig.~\ref{fig:Ay_ppd} are not quantitatively reliable for $\theta_2^\textrm{L} \lesssim 30^\circ$.

%\vspace{-0.5cm}
\begin{figure}[htpb]
 \centering
 \includegraphics[width=0.45\textwidth]{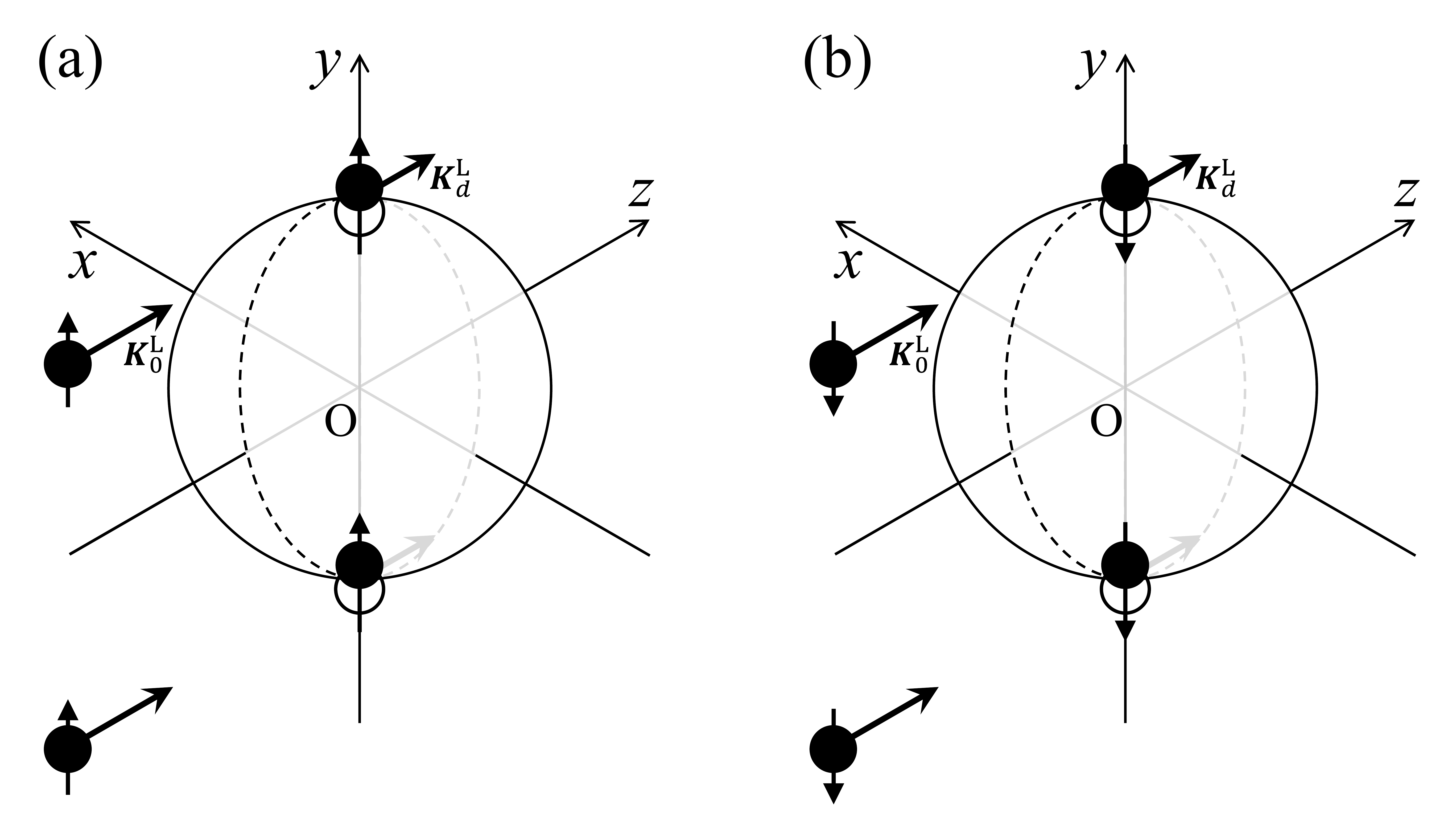}
% \vspace{-0.3cm}
 \caption{Schematic of the initial state for deuteron knockout from orbit $2D_2$.
 In panels (a) and (b), the proton and deuteron spins are oriented upward and downward along the $y$-axis, respectively.
 \label{fig:MarisPol_2D2}}
\end{figure}

The dashed line, a characteristic feature of the ($p,pd$) reaction, lies between the other two lines, and its behavior can be qualitatively explained as follows.
The processes in which a spin-up (spin-down) incident proton knocks out a spin-up (spin-down) deuteron are favored because of the condition $C_{y,y} > 0$, as with the other two orbits.
In addition, by combining the picture in which the deuteron spin is ``orthogonal'' to $\ell$ in orbit $2D_2$ with the condition $K_{d,z}^\trL > 0$, it can be expected that $\ell$ points along the positive (negative) $x$-axis when the deuteron moves counterclockwise (clockwise) in the $y$-$z$ plane shown in Fig.~\ref{fig:MarisPol_2D2}.  % Fig.~\ref{fig:MarisPol}
In this case, the deuteron would be primarily knocked out from regions near the poles of the $y$-axis without strong absorption, indicating that deuterons in both counterclockwise and clockwise motions contribute similarly to the TDX.
Apart from the aforementioned preference, no strong restriction exists on the spin direction of the incident proton.
Therefore, the numerator of Eq.~\eqref{eq:Ay_ppd_y} becomes small, and the $A_y$ for orbit $2D_2$ obtains an intermediate value between those for the other two orbits.
%~\\

%----
\subsection{Nuclear medium effect on the deuteron in the target nucleus\label{sec:NuclMedEff}}

\begin{figure}[htpb]
 \centering
 \includegraphics[width=0.45\textwidth]{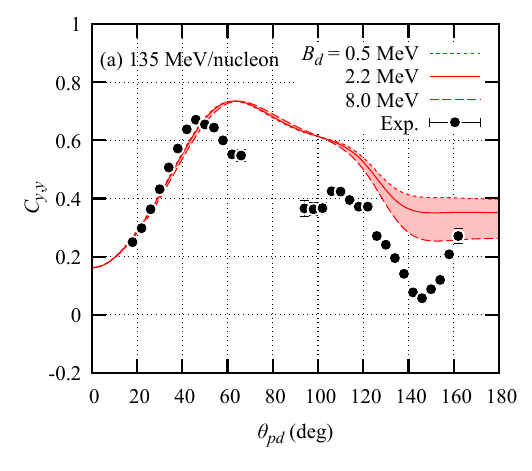}
 \includegraphics[width=0.45\textwidth]{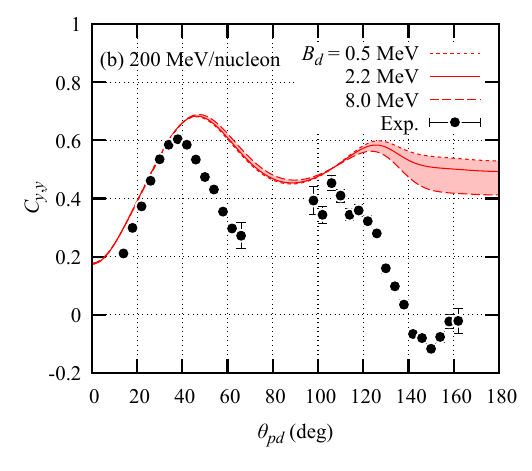}
% \vspace{-0.3cm}
 \caption{Same as Fig.~\ref{fig:Cyy_pd} but the dotted and dashed lines are also shown.
 These lines represent the results of the $d^*$($p,p'$)$d$ processes calculated with $B_d$ values of $0.5$ and $8.0$~MeV, respectively.
 The band indicates the range of variation in $C_{y,y}$ obtained by changing $B_d$ between $0.5$ and $8.0$~MeV.
 \label{fig:Cyy_pd_05-80}}
\end{figure}

\begin{figure}[htpb]
 \centering
 \includegraphics[width=0.45\textwidth]{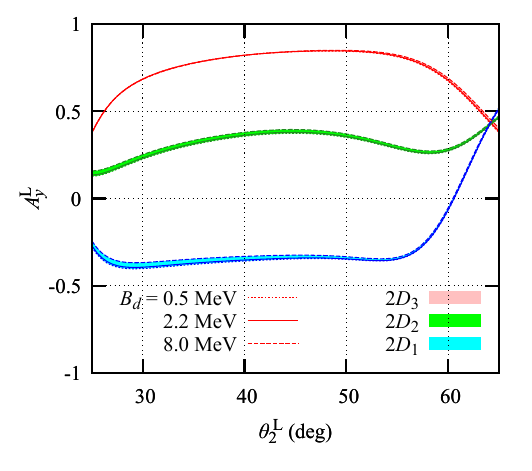}
% \vspace{-0.3cm}
 \caption{Same as Fig.~\ref{fig:Ay_ppd} but the results for different $B_d$ values are also shown.
 The line styles and shaded bands correspond to those in Fig.~\ref{fig:Cyy_pd_05-80}.
 \label{fig:Ay_ppd_05-80}}
\end{figure}

The internal state of the deuteron cluster can differ from that of a free deuteron owing to nuclear medium effects.
In this section, we consider these effects by artificially changing the binding energy $B_d$ of the deuteron in nucleus A from its value in free space, which is approximately $2.2$~MeV.
Because a free deuteron is observed in the final state, a $d^*$($p,p'$)$d$ reaction is considered as the elementary process when $B_d$ differs from $2.2$~MeV, where $d^*$ represents an artificial $p$-$n$ bound state with $B_d \ne 2.2$~MeV.

Figure~\ref{fig:Cyy_pd_05-80} shows the calculated $C_{y,y}$ for various $B_d$ values.
The solid line and filled circles are the same as those in Fig.~\ref{fig:Cyy_pd}.
The dotted and dashed lines represent the results with $B_d = 0.5$ and $8.0$~MeV, respectively, and the band represents the range of $C_{y,y}$ values obtained by varying $B_d$.
The band has a large width at backward angles $\theta_{pd} \gtrsim 120^\circ$, whereas minor differences are observed among the three lines in the angular region between $35^\circ$ and $45^\circ$, which corresponds to the ($p,pd$) kinematics in this study.

Figure~\ref{fig:Ay_ppd_05-80} shows the $A_y$ values of the ($p,pd$) reaction calculated using the same $B_d$ values as those used in the previous paragraph.
For each deuteron-cluster orbit, the line styles correspond to those in Fig.~\ref{fig:Cyy_pd_05-80}.
As expected from the $C_{y,y}$ result, the variation in $A_y$ is small with respect to $B_d$.
Therefore, the conclusions in Secs.~\ref{sec:Cyy_pd} and \ref{sec:56Nippd54Co} would not depend on the internal state of the deuteron cluster for the kinematics considered here.
%~\\

%----
\subsection{Effective interaction dependence\label{sec:EffIntDep}}

\begin{figure}[htpb]
 \centering
 \includegraphics[width=0.45\textwidth]{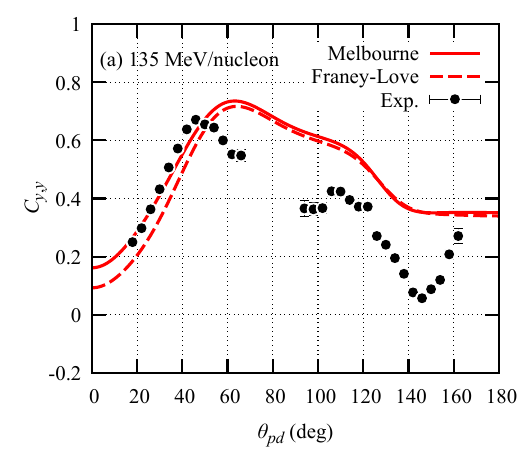}
 \includegraphics[width=0.45\textwidth]{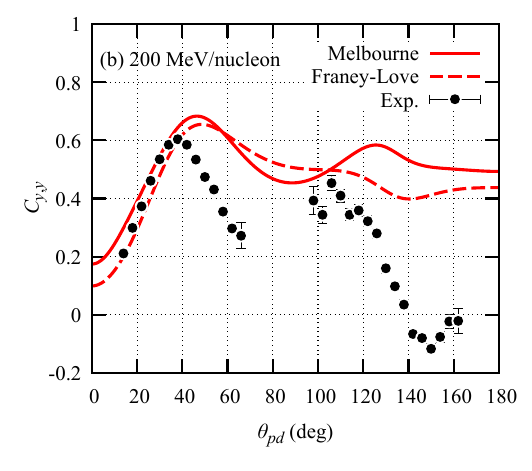}
% \vspace{-0.3cm}
 \caption{Same as Fig.~\ref{fig:Cyy_pd} but the dashed line shows the result calculated using the Franey-Love nucleon-nucleon effective interaction~\cite{MAFraney85}.
 \label{fig:Cyy_pd_FL-Mel}}
\end{figure}

\begin{figure}[htpb]
 \centering
 \includegraphics[width=0.45\textwidth]{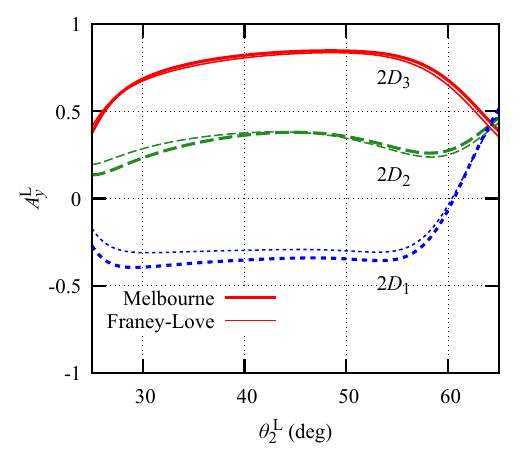}
% \vspace{-0.3cm}
 \caption{Same as Fig.~\ref{fig:Ay_ppd} but the results calculated with the Franey-Love effective interaction are also shown by the thin lines.
 \label{fig:Ay_ppd_FL-Mel}}
\end{figure}

In this section, the dependence of the observables on the nucleon-nucleon effective interaction is discussed.
In Fig.~\ref{fig:Cyy_pd_FL-Mel}, the dashed line represents $C_{y,y}$ of the $p$-$d$ elastic scattering calculated using the Franey-Love effective interaction~\cite{MAFraney85}, whereas the solid and the filled circles are the same as in Fig.~\ref{fig:Cyy_pd}.
The dashed lines show qualitatively similar $\theta_{pd}$ dependences to the solid lines at both energies, reproducing the experimental data well up to approximately $40^\circ$ and overestimating them at angles beyond this value.
However, at angles corresponding to the ($p,pd$) reaction kinematics, the dashed lines are up to approximately $15\%$ and $10\%$ smaller than the solid lines at $135$ and $200$~MeV/nucleon, respectively, indicating that the dominance of spin-parallel processes becomes weaker.

Figure~\ref{fig:Ay_ppd_FL-Mel} shows the comparison of the calculated $A_y$ of the ($p,pd$) reaction with the two effective interactions.
The thick and thin lines correspond to the solid and dashed ones in Fig.~\ref{fig:Cyy_pd_FL-Mel}.
The thin lines differ from the thick ones by up to $20\%$ and move closer to zero, which is consistent with the behavior of $C_{y,y}$.
Nevertheless, the two lines for each orbit exhibit nearly the same $\theta_2^\trL$ dependence, and their signs remain unchanged.
Therefore, the qualitative behaviors of $C_{y,y}$ and $A_y$ shown in Secs.~\ref{sec:Cyy_pd} and \ref{sec:56Nippd54Co} would be similar even for different nucleon-nucleon effective interactions by selecting appropriate kinematics.
%~\\

%----
\section{Summary and perspective\label{sec:Summary}}
The Maris polarization, which is the effective polarization of a particle to be knocked out in a nucleus owing to nuclear absorption and spin-orbit coupling, has been investigated for the ($p,pd$) reaction within the DWIA framework.
Because the deuteron has a spin $1$, three $j$ values are possible for the deuteron-cluster orbits in a target nucleus A with a given $\ell$.
In this study, the $A_y$ value of $^{56}$Ni($p,pd$)$^{54}$Co$^*$ at $250$~MeV was calculated assuming that deuterons were knocked out from orbits $2D_1$, $2D_2$, and $2D_3$, corresponding to $j = 1$, $2$, and $3$, respectively, with $\ell = 2$.
The obtained $A_y$ values for orbits $2D_1$ and $2D_3$ have opposite signs in the considered kinematics.
This behavior is similar to that observed in the ($p,2p$) reaction and indicates that the deuterons in orbits $2D_1$ and $2D_3$ are effectively polarized in the downward and upward directions, respectively.
Regarding orbit $2D_2$, the deuteron would be knocked out from regions near the poles of the $y$-axis because $\ell$ is ``orthogonal'' to the deuteron spin and the deuteron moves in the $y$-$z$ plane in the target nucleus when $K_{d,z}^\trL > 0$.
Weak absorption in these regions results in similar contributions from deuterons in both counterclockwise and clockwise motions, and the TDXs induced by spin-up and spin-down protons largely cancel each other.
Therefore, $A_y$ for orbit $2D_2$ has a smaller magnitude than that for the other orbits.
For the kinematics considered in this research, the conclusion is nearly independent of the deuteron internal state and the effective interaction adopted in the elementary process description.

The results of this work could help to investigate the orbital motion of the deuteron cluster in a nucleus, and the concept of deuteron-cluster orbit may be established.
However, the calculated $A_y$ values have not been compared with experimental data because no such data are found.
To quantitatively understand the Maris polarization in the ($p,pd$) reaction, both experimental and theoretical studies are required for various target nuclei.
A quantitative research on contributions of deuteron breakup effects to $A_y$ is also expected.
%~\\

%----
\begin{acknowledgments}
The author thanks T. Uesaka, K. Ogata, K. Yoshida, and S. Ogawa for fruitful discussions and also thanks Editage for English language editing.
This work was supported in part by Grants-in-Aid for Scientific Research from JSPS (Grant Nos.~JP21H00125 and JP25K17393) and JST ERATO Grant No.~JPMJER2304.
Numerical calculations were performed using the computer facilities at the Research Center for Nuclear Physics, Osaka University.
\end{acknowledgments}

%---- Appendix ----%
%\appendix

%----
%\section{Appendix\label{app:appendix}}

%---- Reference ----%
%\newpage
\bibliography{./References}
%\bibliography{./BibTeX/References}

\end{document}